# Giant third-order nonlinearity amplifies second-harmonic transport in $Ca_3Ru_2O_7$

Subin Mali[1], Yufei Zhao[3], Yu Wang[1], Binghai Yan [1,3] and Zhiqiang Mao [1,2]

[1] Department of Physics, Pennsylvania State University

[2] Department of Materials Science and Engineering, Pennsylvania State University

[3] Department of Condensed Matter Physics, Weizmann Institute of Science

## Abstract

Nonlinear electrical transport enables the conversion and detection of electromagnetic signals, supporting wireless energy harvesting, sensing and frequency conversion. Achieving both a large response and electrical tunability remains a central materials challenge. Here we harness giant third-order nonlinearity to amplify and electrically control second-harmonic transport in the correlated polar metal $Ca_3Ru_2O_7$. Superposing a dc bias on an ac excitation enhances the second-harmonic signal by more than an order of magnitude. The bias controls the polarity of the induced response and activates channels forbidden by symmetry at zero bias. Both transverse and longitudinal components scale quadratically with the ac excitation and are odd in the dc bias, with angular dependences consistent with third-order conductivity. The underlying third-harmonic electric-field responsivity reaches $4.5\times10^{-6}$ $m^2V^{-2}$ at 3 K, providing a quantitative measure of the strong nonlinearity. The induced response emerges below the spin-reorientation transition near 48 K, linking its enhancement to the reconstructed low-temperature electronic state. Theoretical calculations identify quantum-geometric contributions to the nonlinear conductivity and reproduce the observed response symmetry. These results demonstrate how strong higher-order nonlinearity can amplify an effective second-order response, providing a route to electrically tunable nonlinear transport in correlated materials.

## Introduction

Nonlinear electrical transport has emerged as a powerful probe of quantum geometry, providing access to information that is inaccessible to conventional linear-response measurements [1–3]. Whereas linear conductivity is governed primarily by carrier density and scattering, nonlinear responses can be sensitive to crystal symmetry and geometric properties of Bloch wave functions. This has stimulated extensive studies of second-order nonlinear transport. The Berry-curvature-dipole-driven nonlinear Hall effect, first proposed for time-reversal-invariant inversion-broken materials [4,5] and experimentally observed in bilayer [6] and few-layer $WTe_2$ [7], has since been reported in a variety of low-symmetry materials and symmetry-engineered systems, including type-II Weyl semimetals such as $TaIrTe_4$ [8], $T_d$–$MoTe_2$ and $WTe_2$ [9], corrugated bilayer graphene [10], moiré systems such as twisted double-bilayer graphene [11] and twisted bilayer $WSe_2$ [12], bulk materials such as $BaMnSb_2$ [13] and $NbIrTe_4$ [14]. Second-harmonic Hall responses arising from extrinsic mechanisms, including skew scattering and side jump, have also been reported in topological-insulator surface states such as $Bi_2Se_3$ [15], and magnetic heterostructure [16]. More recently, the quantum metric dipole [17–19], the symmetric counterpart to Berry curvature in the quantum geometric tensor, has been shown to drive second-order transport in few-layer antiferromagnetic topological insulator $MnBi_2Te_4$ [20,21]. Together, these advances establish second-order nonlinear transport as a sensitive probe of quantum geometry and highlight its potential for rectification, frequency conversion, and wireless sensing. However, nonlinear signals are often intrinsically weak and strongly constrained by symmetry and material-specific band geometry, making large, robust, and electrically tunable responses difficult to realize.

Higher-order nonlinear transport offers a route beyond these limitations. Third-order responses are subject to fewer symmetry constraints than second-order responses and have recently emerged as probes of higher moments of quantum geometry. Third-order responses have been reported in FeSn thin films [22], $T_d$-$MoTe_2$ [23], $Fe_5GeTe_2$ [24], $MnBi_2Te_4$ [25], and bulk $TaIrTe_4$ [26], where they were linked to the Berry-curvature quadrupole, Berry-connection polarizability and quantum metric quadrupole. Importantly, third-order conductivity enables mixing a static dc electric field with an ac drive to generate an effective second-harmonic response, providing direct electrical control of its magnitude and sign [27,28]. To date, dc-field-driven second-harmonic transport has been demonstrated primarily in weakly correlated Weyl semimetals [27–29]. Whether strong electronic correlations can substantially amplify this field-induced nonlinear response remains an open question.

Growing evidence suggests strong electronic correlations can amplify nonlinear responses [30]. In the Weyl-Kondo semimetal candidate $Ce_3Bi_4Pd_3$ the measured nonlinear Hall response was reported to exceed the corresponding first-principles estimate by nearly three orders of magnitude [31], pointing to an important role for correlation beyond a noninteracting band description. More generally, coupling among charge, spin, orbital, and lattice degrees of freedom can reconstruct and renormalize the low-energy electronic structure, giving rise to narrow bands, reconstructed Fermi pockets, and avoided crossings that can generate sharp momentum-space variations of Berry curvature and quantum metric. Because higher-order nonlinear conductivities depend sensitively on the detailed band dispersion and on momentum-space variations of quantum-geometric quantities, they can be particularly sensitive to such reconstructed band features. Therefore, a correlated material hosting a reconstructed topological electronic state may exhibit a nonlinear response far larger than expected from symmetry considerations alone.

$Ca_3Ru_2O_7$ (CRO) provides a promising platform to investigate the role of electronic correlations and band reconstruction in enhancing third-order nonlinear transport. It is a correlated polar metal in which spin, lattice, orbital, and electronic degrees of freedom are strongly coupled. It exhibits an antiferromagnetic transition at $T_N \approx 56$ K with magnetic moments along the *a*-axis (AFM-*a*), followed by a spin-reorientation transition at $T_S \approx 48$ K to the AFM-*b* state, accompanied by a pronounced reconstruction of the low-energy electronic structure [32,33]. This material also displays a rich variety of emergent phenomena, including colossal magnetoresistance [34,35], spin-valve effects [36], metamagnetic textures [37,38], and polar semimetal behavior [39,40]. Angle-resolved photoemission spectroscopy revealed a $C_2$-symmetric reconstructed low energy state with massive Dirac-like bands and identified the spin-reorientation transition as electronically driven in this correlated polar metal [41]. More recent theoretical studies suggest that the linear bands observed by ARPES may instead originate from a Weyl chain semimetal state [42]. The low-temperature crystal structure of CRO remains debated, with $Bb2_1m$ and $Pn2_1a$ proposed as candidate non-centrosymmetric structures [42–44]. In our recent work, second-order nonlinear transport distinguished the symmetry-allowed conductivity tensors of these structures and supported the $Pn2_1a$ symmetry below $T_S$ [45]. However, the absence of diffraction signatures unique to the $Pn2_1a$ structure suggests that the observed symmetry lowering may be predominantly electronic in origin and driven by correlations [46]. The combination of strong correlations, reconstructed topological bands, and low symmetry therefore makes CRO a particularly favorable platform for investigating correlation-enhanced higher-order nonlinear transport.

Here, we demonstrate dc-field-driven second-harmonic transport in $Ca_3Ru_2O_7$ arising from its third-order nonlinear conductivity. Superposing a dc bias with an ac excitation enhances the second-harmonic voltage response by up to a factor of 25. The induced signal is quadratic in the ac current, odd in the dc bias, and exhibits the angular dependence expected from the third-order conductivity tensor. Crucially, the response vanishes above the spin-reorientation transition at $T_S \sim 48$ K and emerges sharply in the reconstructed low-temperature phase, linking the large nonlinear response to correlation-driven electronic reconstruction. More broadly, these results demonstrate that, while symmetry determines whether field-driven nonlinear transport is allowed, its magnitude can be strongly enhanced by a reconstructed electronic structure with pronounced quantum-geometric features. These findings establish correlated topological materials as promising platforms for large, electrically tunable nonlinear responses and point toward new opportunities for nonlinear-transport-based technologies.

**Results**

To investigate dc-field-enhanced nonlinear transport in $Ca_3Ru_2O_7$, we fabricated disk-shaped microdevices from single-crystal flakes using photolithography and focused-ion-beam (FIB), as detailed in Supplementary Material [47]. A representative device consists of twelve electrodes contacting a ~1 μm-thick CRO flake (Fig. 1a). In-plane angle-dependent magnetoresistance measurements confirm that the flake is predominantly single-domain (Fig. 1b). This geometry enables large in-plane electric fields over short channel lengths, avoids complications of multiple domains [48], and allows independent control of the ac current and the dc-bias direction in the *ab*-plane. We determined the crystallographic axes from the in-plane magnetoresistance. For $B \parallel b$, a sharp spin-valve transition near 6 T marks the field-driven transition from the AFM-b state to a canted antiferromagnetic (CAFM) state and is accompanied by a pronounced drop in resistivity, whereas no comparable abrupt transition occurs for $B \parallel a$, consistent with previous reports [35,36]. The step-like switching observed during in-plane field rotation was used to calibrate the $a$- and $b$-axis orientations. Figure 1c shows the temperature-dependent resistivity together with the dc-bias-induced second-harmonic response. The resistivity exhibits clear anomalies at the Néel temperature $T_N \approx 56$ K and the spin-reorientation temperature $T_S \approx 48$ K, consistent with previous studies [40,49,50].

**Table 1** Symmetry-permitted and symmetry-forbidden components of the third-order nonlinear conductivity tensor $\sigma_{i;jkl}$ for in-plane electric fields. Here, the first index $i$ denotes the direction of the nonlinear current response, while $j$, $k$, and $l$ denote the directions of the applied electric fields. The permitted components can contribute to the dc-field-induced second-harmonic response $j_i^{2\omega} = \sigma_{i;jkl}^{(3)} E_j^{\omega} E_k^{\omega} E_l^{\mathrm{dc}}$, whereas the forbidden components vanish by symmetry.

| | Symmetry permitted | Symmetry forbidden |
|---|---|---|
| Third-order tensors | $\sigma_{a;aaa}$, $\sigma_{a;bba}$, $\sigma_{b;aab}$, $\sigma_{b;bbb}$ | $\sigma_{a;bbb}$, $\sigma_{a;aab}$, $\sigma_{b;bba}$, $\sigma_{b;aaa}$ |

The $Bb2_1m$ and $Pn2_1a$ structures impose the same symmetry selection rules on the in-plane third-order nonlinear response because they share the same relevant symmetry operations $\left\{C_{2y}\middle|\left(0,\frac{b}{2},0\right)\right\}$, $\widetilde{M}_x = \left\{M_x\middle|\left(\frac{a}{2},\frac{b}{2},\frac{c}{2}\right)\right\}$, and $\widetilde{M}_z = \left\{M_z\middle|\left(\frac{a}{2},0,\frac{c}{2}\right)\right\}$. Consequently, the same third-order conductivity components are allowed or forbidden in both structures. Thus, unlike the second-order response, the in-plane third-order response does not distinguish between these two candidate structures. The symmetry-allowed and symmetry-forbidden components of the third-order nonlinear conductivity tensor are summarized in Table 1. The mirror symmetry $\widetilde{M}_x$ would give rise to an angle dependence on the nonlinear response. With the applied dc electric field $E^{dc} = (E_\theta^{dc}\sin\theta, E_\theta^{dc}\cos\theta)$ within *ab* plane, one can describe the third-order current density as

$$j_a^{2\omega} = E_a^2 E_\theta^{dc}\,\sigma_{a;aaa}\sin\theta + E_b^2 E_\theta^{dc}\sigma_{a;bba}\sin\theta, \tag{1}$$

$$j_b^{2\omega} = E_a^2\sigma_{b;aa} + E_b^2\sigma_{b;bb} + E_a^2 E_\theta^{dc}\,\sigma_{b;aab}\cos\theta + E_b^2 E_\theta^{dc}\sigma_{b;bbb}\cos\theta. \tag{2}$$

Here $\sigma_{b;aa}$ and $\sigma_{b;bb}$ represent the second-order conductivity, which is $I_{dc}$ - independent. $\theta$ is defined as the angle between the electric field and the *b*-axis.

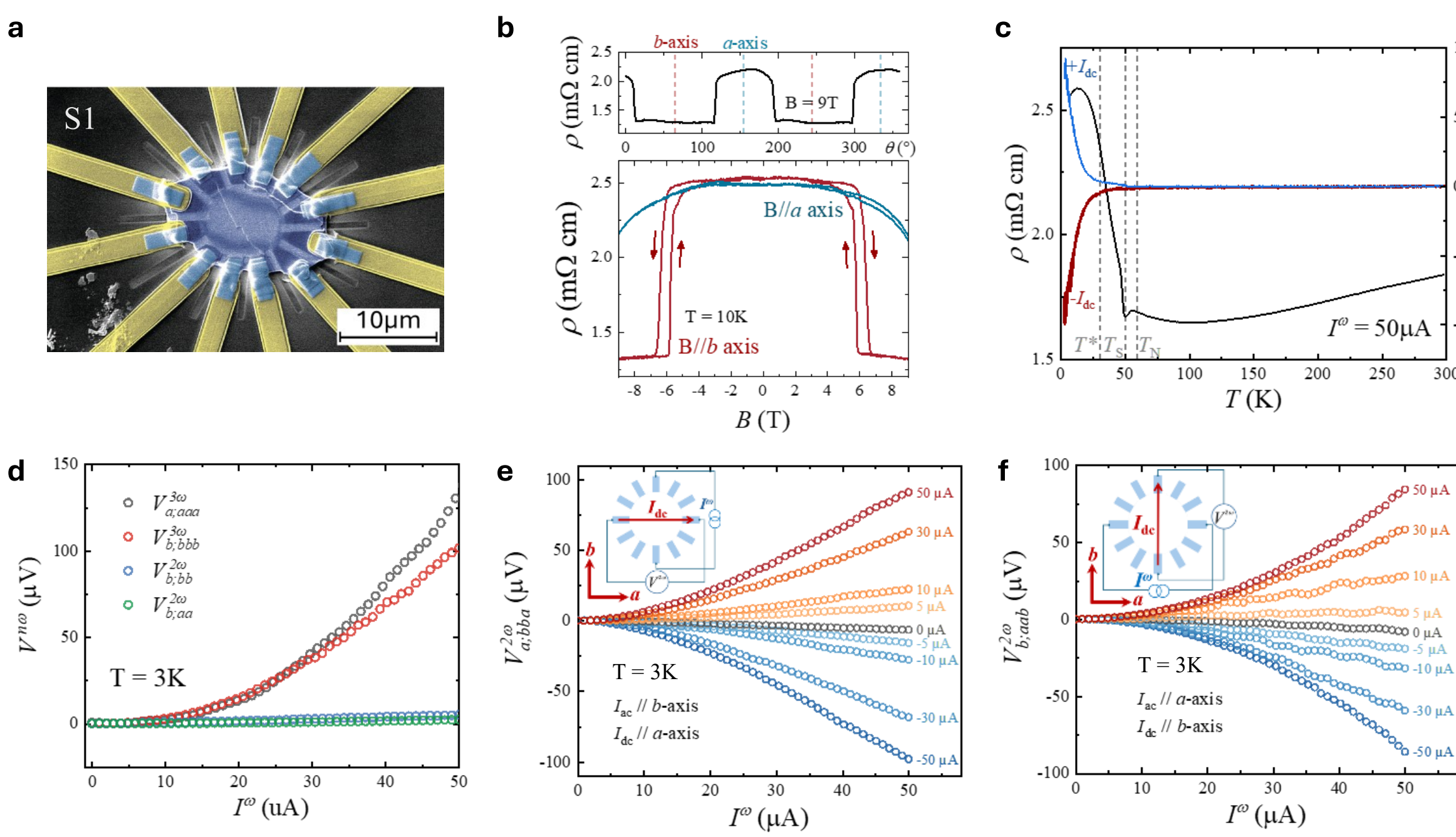

**Figure 1. Device characterization and dc-field-driven second-order Hall response for CRO. (a)** False-colored scanning electron microscope image of device S1 with 12 electrodes. **(b)** In-plane field-orientation-dependent resistivity (upper panel) and field dependent in-plane resistivity for $B \parallel b$ and $B \parallel a$ (bottom panel). **(c)** Temperature

dependence of the linear resistivity and dc-bias-induced second-harmonic Hall response. Vertical dashed lines mark the characteristic temperatures $T_N$ (Neel temperature ~56K), $T_S$ (spin reorientation transition temperature ~ 48 K), and $T^*$ (~ 30K, below which the intrinsic 2nd order nonlinear response enhances sharply [45]). **(d)** Second and third harmonic voltage as a function of ac excitation current $I^{\omega}$ at $T = 3$K for the symmetry-allowed nonlinear transport channels. **(e)** Second-harmonic transverse voltage $V_{a;bba}^{2\omega}$ measured as a function of ac current $I^{\omega}$ for different dc bias currents $I_{\mathrm{dc}}$, with $I^{\omega} \parallel b$-axis and $I_{\mathrm{dc}} \parallel a$-axis. **(f)** Second-harmonic transverse voltage $V_{b;aab}^{2\omega}$ measured with $I^{\omega} \parallel a$-axis and $I_{\mathrm{dc}} \parallel b$-axis. Inset in **(e)** and **(f)** show the schematic illustration of the experimental configurations.

**DC field-induced nonlinear Hall response**

We first examined the dc-field-induced transverse second-harmonic response in two orthogonal configurations. In Fig. 1e, an ac current $I^{\omega}$ was applied along crystallographic *b*-axis, while a dc current $I_{\mathrm{dc}}$ was applied along the orthogonal $a$-axis, and the transverse second-harmonic voltage $V_{a;bba}^{2\omega}$ was detected along the *a*-axis. At zero dc bias, $V_{a;bba}^{2\omega}$ is negligible, consistent with the symmetry-forbidden second-order tensor component $\sigma_{a;bb}^{(2)}$ [42,45]. A finite dc bias, however, generates a sizable $V_{a;bba}^{2\omega}$ that scales approximately as $(I^{\omega})^2$ and reverses sign with $I_{\mathrm{dc}}$, consistent with a third-order nonlinear current of the form $j_i^{2\omega} = \sigma_{i;jkl}^{(3)} E_j^{\omega} E_k^{\omega} E_l^{dc}$. At $I_{\mathrm{dc}}$ = 50 μA, $V_{a;bba}^{2\omega}$ reaches approximately 100 μV, about 20 times larger than the purely ac-driven second-order response under comparable drive conditions (Fig. 1d).

The reciprocal configuration shows the same behavior, as shown in Fig. 1f. With $I^{\omega}$ along the *a*-axis, $I_{\mathrm{DC}}$ along the *b*-axis, and $V_{b;aab}^{2\omega}$ detected along the *b*-axis, a small finite signal of ~ 5 μV remains at zero dc bias, consistent with the non-zero second-order tensor component $\sigma_{b;aa}^{(2)}$ [42,45]. Applying $I_{\mathrm{dc}}$ strongly enhances $V_{b;aab}^{2\omega}$ to approximate 85 μV at $I_{\mathrm{dc}}$ = 50 μA, corresponding to a 17-fold increase. The response is likewise approximately quadratic in $I^{\omega}$ and odd under reversal of $I_{\mathrm{dc}}$, consistent with the same dc-mixed third-order origin.

The unusually large third-order response is also evident from the direct comparison in Fig. 1d. Under purely ac excitation, the third-harmonic voltages $V_{a,aaa}^{3\omega}$and $V_{b,bbb}^{3\omega}$ reach approximately 100-125 μV at 50 μA, about 20–25 times larger than the intrinsic second-harmonic signals ($V_{a,bb}^{2\omega}$ and $V_{b,bb}^{2\omega}$) of approximately 2–5 μV. This is remarkable because, in the perturbative regime, higher-order nonlinear responses are generally expected to be weaker than lower-order ones. The temperature dependence in Fig. 1c further reveals that the dc-mixed second-harmonic response arising from this large third-order conductivity is intimately linked to the reconstructed low-temperature phase. The signal is negligible above $T_{\mathrm{S}}$, emerges below the spin-reorientation transition, and grows sharply below $T^* \approx 30$ K, coincident with the strong enhancement of the purely ac-driven second-order response [45]. Moreover, reversing the dc-bias polarity reverses the second-harmonic signal throughout the low-temperature regime where the response is observed. Together, these observations show that a static in-plane electric field mixes with two ac fields through the third-order conductivity to generate a large, electrically tunable second-harmonic response, whose onset is closely tied to the low-temperature electronic reconstruction of CRO [41].

To resolve the tensor structure of the dc-field-induced nonlinear response, we rotated $I_{dc}$ within the *ab* plane while keeping the ac-current and voltage-detection directions fixed (Figs. 2a and 2d). In both transverse configurations, the induced second-harmonic voltage is approximately linear with dc bias at low currents, with a slope that depends strongly on the dc-current orientation (Figs. 2b and 2e). At larger dc bias, the response becomes sublinear, indicating a deviation from the lowest-order perturbative behavior. The signals are strongly suppressed when the dc field approaches a symmetry-forbidden orientation, while small residual signals may arise from device misalignment, nonideal circular geometry, or contact asymmetry. Figure 2c and 2f summarize the angular dependence of the field-normalized second-harmonic response $E^{2\omega}_{b;aa\theta}/E^3$ and $E^{2\omega}_{a;bb\theta}/E^3$, where $E^3 \equiv (E^{\omega})^2 E^{\mathrm{dc}}_{\theta}$ represents the product of two ac electric fields and one dc electric field. For an ac excitation along the $a$-axis, Eq. (2) gives $j^{2\omega}_b = E^2_a \sigma_{b;aa} + E^2_a E^{dc}_{\theta}\ \sigma_{b;aab}\cos\theta$, whereas for an ac excitation along the $b$-axis, Eq (1) gives $j^{2\omega}_a = E^2_b E^{dc}_{\theta} \sigma_{a;bba}\sin\theta$. Since the intrinsic second-order contribution is comparatively small (Fig. 1d), the dc-induced response is dominated by the third-order term. Accordingly, the normalized responses follow complementary cosine (Fig. 2c) and sine (Fig. 2f) dependences, consistent with the symmetry-allowed tensor components $\sigma_{b;aab}$ and $\sigma_{a;bba}$, respectively. The dc-induced signals also reverse sign upon reversal of $I_{dc}$, supporting their third-order origin. Together, these measurements demonstrate that the dc field selectively accesses and controls different components of the third-order nonlinear conductivity tensor, rather than simply amplifying the intrinsic second-order response.

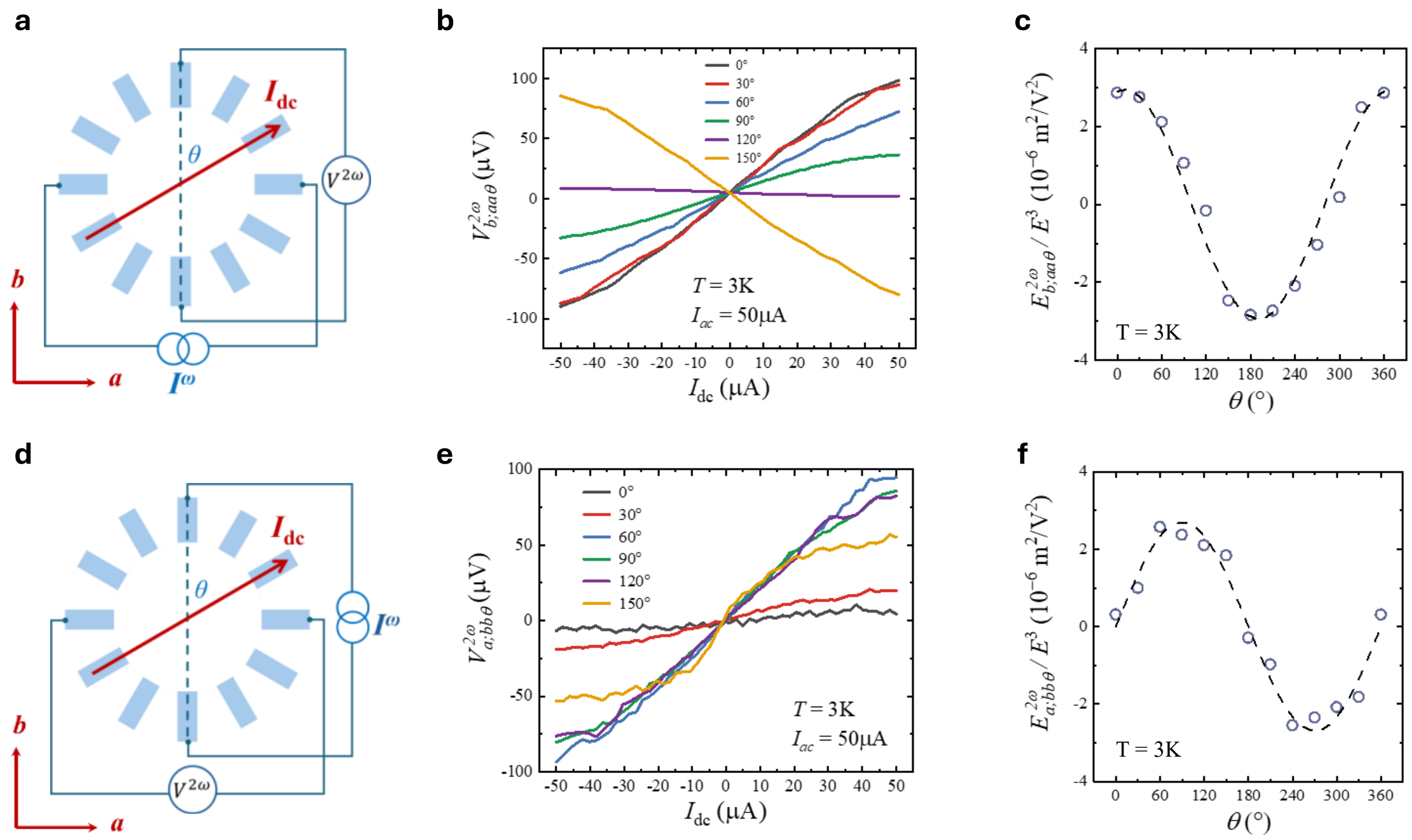

**Figure 2. Angular dependence of the dc-field-induced second-order Hall response. (a)** Schematic of the measurement configuration for $V^{2\omega}_{b;aa\theta}$. An ac current $I^{\omega}$is applied along the $a$-axis, a dc bias current $I_{\mathrm{dc}}$ is applied at an in-plane angle $\theta$ relative to the $b$-axis, and the transverse second-harmonic voltage is measured along the $b$-axis. **(b)** $V^{2\omega}_{b;aa\theta}$ as a function of $I_{\mathrm{dc}}$ for different dc-bias angles $\theta$, measured at fixed $I^{\omega} = 50\ \mu$A. The response is odd in $I_{\mathrm{dc}}$ and changes systematically with $\theta$. **(c)** Angular dependence of the normalized nonlinear response $E^{2\omega}_{b;aa\theta}/E^3$. The dashed curve denotes the sinusoidal fit to the experimental data. **(d)** Schematic for the measurement configuration for $V^{2\omega}_{a;bb\theta}$, where $I^{\omega}$ is applied along the $b$-axis, $I_{\mathrm{dc}}$ is rotated in the $ab$-plane, and the transverse second-harmonic voltage is measured along the $a$-axis. **(e)** $V^{2\omega}_{a;bb\theta}$ as a function of $I_{\mathrm{dc}}$ for different $\theta$, measured at fixed $I^{\omega} = 50\ \mu$A. (**f)** Angular dependence of the normalized response $E^{2\omega}_{a;bb\theta}/E^3$, extracted from the slopes in (**e**). The dashed curve shows the corresponding sinusoidal fit.

### DC field-induced nonlinear longitudinal response

We next examined the dc-field-induced longitudinal second-harmonic response, with the ac current and voltage detection along the same crystallographic axis. Figure 3a and 3d shows $V^{2\omega}_{a;aaa}$ and $V^{2\omega}_{b;bbb}$ measured as a function of $I^{\omega}$ for different dc-bias currents applied along the $a$- and $b$-axes respectively. Similar to the transverse responses, both $V^{2\omega}_{a;aaa}$ and $V^{2\omega}_{b;bbb}$ are strongly enhanced by the dc bias, reverses sign upon reversal of $I_{\mathrm{dc}}$, and remains approximately quadratic in $I^{\omega}$. These characteristics are consistent with a mixed third-order response involving two ac fields and one dc field.

To further test the tensor character of the longitudinal response, we rotated the dc current within the $ab$-plane while keeping the ac-current and voltage-detection directions fixed. In both longitudinal configurations (Fig. 3b & 3e), the dc-induced second harmonic responses varies approximately linearly with $I_{\mathrm{dc}}$ at low bias, with a strongly angle-dependent slope. The normalized responses $E^{2\omega}_{a;aa\theta}/E^3$ and $E^{2\omega}_{b;bb\theta}/E^3$ follow complementary sine (Fig. 3c) and cosine (Fig. 3f) dependences, respectively, consistent with the symmetry-allowed tensor components $\sigma_{a;aaa}$ and $\sigma_{b;bbb}$ described by Eqs. (1) and (2). The strong suppression of the response near symmetry-forbidden dc-field orientations further supports its tensor-governed origin and argues against contact asymmetry or current-even thermal backgrounds. Together with the transverse measurements, these results establish that the applied dc field selectively accesses and controls symmetry-allowed components of the third-order nonlinear conductivity tensor.

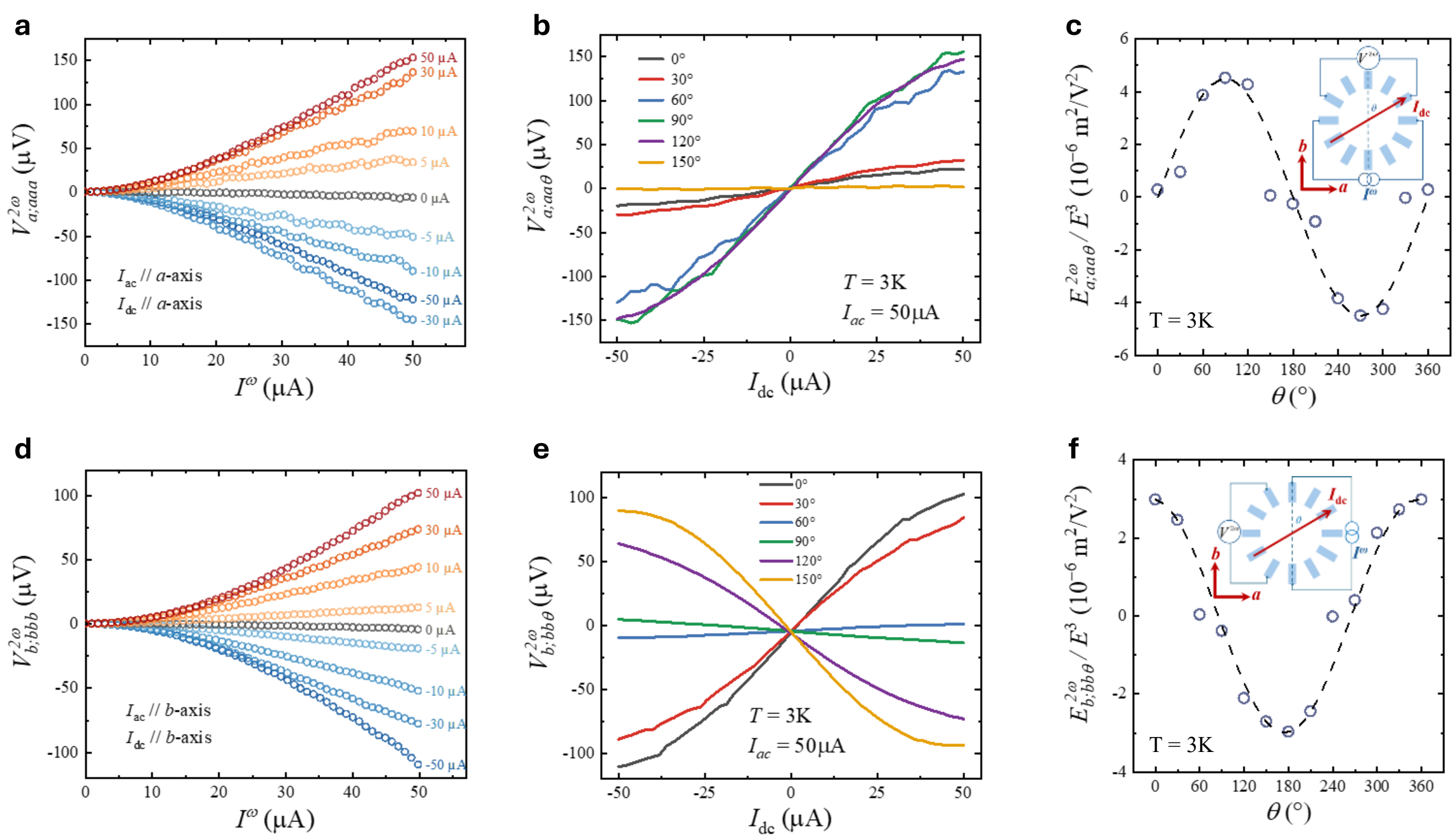

**Figure 3. DC-field-induced longitudinal second-harmonic response. (a)** Longitudinal second-harmonic voltage $V^{2\omega}_{a;aaa}$ as a function of ac current $I^{\omega}$ for different $I_{\mathrm{dc}}$, with both currents applied along the $a$-axis. **(b)** $V^{2\omega}_{a;aa\theta}$ as a function of $I_{\mathrm{dc}}$ for different in-plane dc-current angles $\theta$ at fixed $I^{\omega}$=50 $\mu$A. **(c)** Angular dependence of the normalized nonlinear response $E^{2\omega}_{a;aa\theta}/E^3$. The dashed curve denotes the expected sinusoidal dependence; the inset illustrates the measurement configuration. **(d)** Longitudinal second-harmonic voltage $V^{2\omega}_{b;bbb}$ as a function of $I^{\omega}$ for different $I_{\mathrm{dc}}$, with both currents along the $b$-axis. **(e)** $V^{2\omega}_{b;bb\theta}$ as a function of $I_{\mathrm{dc}}$ for different dc-bias angles $\theta$, at fixed $I^{\omega} = 50$ $\mu$A. **(f)** Angular dependence of the normalized response $E^{2\omega}_{b;bb\theta}/E^3$. The dashed curve denotes the expected sinusoidal dependence; the inset illustrates the measurement configuration.

## Discussion

Our measurements establish the dc-bias-induced second-harmonic response in CRO as a manifestation of third-order nonlinear conductivity in the reconstructed low-temperature phase. Both transverse (Fig. 1 & 2) and longitudinal signals (Fig. 3) scale quadratically with the ac excitation, are odd in the applied dc field, and exhibit the angular dependence (Fig. 2c & 2f and Fig. 3c & 3f) expected from the third-order conductivity tensor. The underlying physics can be understood by comparing the conventional third-harmonic response,

$$j_i^{3\omega} = \sigma_{i;jkl}^{(3)} E_j^{\omega} E_k^{\omega} E_l^{\omega},$$

with the dc-field-induced second-harmonic response,

$$j_i^{2\omega} = \sigma_{i;jkl}^{(3)} E_j^{\omega} E_k^{\omega} E_l^{\mathrm{dc}}.$$

Thus, replacing one ac-field input with a dc field converts the third-order conductivity into an effective second-harmonic response. The dc field therefore enables direct electrical control over both the magnitude and sign of the nonlinear response. Importantly, this field-mixing mechanism not only enhances the second-harmonic response but also enables second-harmonic transport in geometries where the intrinsic second-order response is symmetry-forbidden.

Our experiments also show that the dc-field-induced 2$^{\mathrm{nd}}$ harmonic response is associated with the intrinsic low-energy electronic state of CRO rather than a device-specific artifact. First, the same characteristic behavior is reproduced in two independently fabricated devices: quadratic scaling with ac-current amplitude, odd dependence on dc-bias polarity, approximately linear dependence on dc bias in the weak-field regime, and symmetry-governed angular variation (see Supplementary Note 2 for details on the other device). Second, the nonlinear response is negligible above the spin-reorientation transition at $T_{\mathrm{S}}$ and becomes pronounced upon entering the AFM-*b* phase. The reproducible onset near the same bulk transition in different devices strongly links the response to the reconstructed electronic state of CRO. In addition, the systematic suppression of the signal when the dc field is rotated toward symmetry-forbidden directions is difficult to reconcile with simple contact rectification or current-even thermal backgrounds. These observations strongly support an electronic origin of the dc-field-induced nonlinear response.

CRO also illustrates that symmetry allowance alone is insufficient to account for a large third-order response. Because third-order nonlinear conductivity is subject to fewer symmetry restrictions than second-order conductivity, both the high- and low-temperature phases of CRO can in principle support third-order transport. The sharp enhancement below $T_{\mathrm{S}}$ therefore points to an additional mechanism beyond symmetry. Moreover, in the perturbative regime, higher-order responses are generally expected to be weaker than lower-order ones, whereas in CRO the third-harmonic signal exceeds the intrinsic second-harmonic response by approximately 20–25 times (Fig. 1d) at around 50 μA current. We attribute this enhancement to the reconstructed low-energy electronic state below $T_S$, where strong correlations, small Fermi pockets, and topological band features can strongly amplify the sensitivity of higher-order transport to states near the Fermi level. Consistent with this picture, CRO exhibits the largest third-order nonlinear responsivity among the representative materials summarized in Supplementary Table 1.

Electronic correlations offer a possible mechanism for this enhancement. Previous second-order transport measurements supported the electronic symmetry of $Pn2_1a$, whereas synchrotron diffraction did not resolve reflections unique to that structure [45,46], suggesting that the symmetry reduction below $T_S$ is predominantly electronic rather than arising from a sizable static structural distortion. Such electronically driven symmetry lowering provides further evidence for a correlation-induced reconstruction of the low-energy electronic state and naturally links the emergence of the giant third-order nonlinear response to the reconstructed phase below $T_S$. This interpretation is consistent with growing evidence that electronic correlations can strongly modify nonlinear response. In weakly correlated systems, the nonlinear response is largely governed by the single-particle band structure and the associated band-geometric quantities near the Fermi level. By contrast, in correlated materials, many-body interactions can renormalize and reconstruct the low-energy electronic structure, potentially amplifying nonlinear responses. This possibility is supported by studies of the Weyl–Kondo semimetal $Ce_3Bi_4Pd_3$, where a giant nonlinear Hall response was observed experimentally and theoretical calculations showed that strong correlations can substantially enhance both nonlinear longitudinal and Hall conductivities [30,31]. Enhanced optical higher order harmonic generation has been observed experimentally in correlated electron systems and investigated theoretically [51–54], indicating that many-body interactions can substantially modify nonlinear optical and transport coefficients. More recently, the large nonlinear Hall response in few-layer $TaIrTe_4$ was found to be associated with a correlated charge-density-wave state that reconstructs the low-energy bands and redistributes Berry curvature [55]. These examples support the broader view that correlated electronic reconstruction can strongly enhance nonlinear transport by reshaping the band structure and band-geometric quantities near the Fermi level.

Our first-principles calculations provide additional support for this interpretation. The calculations reproduce the symmetry structure of the observed third-order response and identify contributions from the Drude, Berry-curvature-quadrupole, quantum-metric-quadrupole, and symplectic-connection terms (see Supplementary Material [47]). However, the calculated third-order conductivity remains substantially smaller than the experimentally inferred response. The discrepancy indicates that the magnitude of the nonlinear conductivity is not fully captured by the present DFT+$U$ treatment. Given the known correlation-driven reconstruction of CRO, additional renormalization of the low-energy quasiparticles provides a plausible source of enhancement. Changes in quasiparticle mass, scattering rate, orbital hybridization, chemical potential, and the positions of small Fermi pockets can all strongly affect a higher-order response and may contribute to the difference between calculation and experiment.

Among the materials summarized in Supplementary Table 1, the altermagnetic candidate $RuO_2$ exhibits the strongest previously reported third-order nonlinear response and therefore provides a natural benchmark for CRO. Extracting the bulk third-order conductivity of $RuO_2$ from Ref [56], we obtain $\sigma^{(3)}_{\mathrm{RuO_2}} \approx 5 \times 10^{-2}\ \mathrm{m\,\Omega^{-1}V^{-2}}$ compared with $\sigma^{(3)}_{\mathrm{CRO}} \approx 9 \times 10^{-1}\ \mathrm{m\,\Omega^{-1}V^{-2}}$ extracted from our measurements. Thus, the third-order conductivity of CRO is approximately 18 times larger than that of $RuO_2$, highlighting its exceptionally strong bulk nonlinear response. The large responses of CRO and $RuO_2$, both Ru-based oxides with spin-orbit-coupled $4d$ electronic states, motivate further investigation of whether Ru-derived electronic states provide favorable conditions for enhanced third-order transport. More broadly, these results identify correlated topological materials with tunable electronic reconstructions as promising platforms for realizing giant, electrically controlled nonlinear transport.

**Summary**

In summary, we demonstrate a giant dc-field-driven second-harmonic response in $Ca_3Ru_2O_7$ arising from its third-order nonlinear conductivity. The response is quadratic in the ac excitation, odd in the dc bias, and follows the symmetry of the third-order conductivity tensor. Its emergence below $T_S \approx 48\mathrm{K}$, together with the unusually large third-order response, points to a strong enhancement associated with the reconstructed correlated electronic state. These results establish correlated topological materials as promising platforms for large, electrically tunable nonlinear transport.

## Acknowledgments

This work is primarily supported by the US National Science Foundation under grant DMR 2211327.

**Reference:**

[1] Y. Jiang, T. Holder, and B. Yan, Revealing quantum geometry in nonlinear quantum materials, Reports on Progress in Physics **88**, 076502 (2025).

[2] A. Gao, N. Nagaosa, N. Ni, and S.-Y. Xu, Quantum geometry phenomena in condensed matter systems, Rev. Mod. Phys. **98**, 035005 (2026).

[3] Z. Z. Du, H.-Z. Lu, and X. C. Xie, Nonlinear Hall effects, Nat Rev Phys **3**, 744 (2021).

[4] I. Sodemann and L. Fu, Quantum Nonlinear Hall Effect Induced by Berry Curvature Dipole in Time-Reversal Invariant Materials, Phys. Rev. Lett. **115**, 216806 (2015).

[5] Y. Zhang, Y. Sun, and B. Yan, Berry curvature dipole in Weyl semimetal materials: An ab initio study, Phys. Rev. B **97**, 041101 (2018).

[6] Q. Ma et al., Observation of the nonlinear Hall effect under time-reversal-symmetric conditions, Nature **565**, 337 (2019).

[7] K. Kang, T. Li, E. Sohn, J. Shan, and K. F. Mak, Nonlinear anomalous Hall effect in few-layer $WTe_2$, Nature Materials **18**, 324 (2019).

[8] D. Kumar, C.-H. Hsu, R. Sharma, T.-R. Chang, P. Yu, J. Wang, G. Eda, G. Liang, and H. Yang, Room-temperature nonlinear Hall effect and wireless radiofrequency rectification in Weyl semimetal $TaIrTe_4$, Nature Nanotechnology **16**, 421 (2021).

[9] A. Tiwari et al., Giant c-axis nonlinear anomalous Hall effect in $T_d$-$MoTe_2$ and $WTe_2$, Nature Communications **12**, 2049 (2021).

[10] S.-C. Ho, C.-H. Chang, Y.-C. Hsieh, S.-T. Lo, B. Huang, T.-H.-Y. Vu, C. Ortix, and T.-M. Chen, Hall effects in artificially corrugated bilayer graphene without breaking time-reversal symmetry, Nature Electronics **4**, 116 (2021).

[11] S. Sinha et al., Berry curvature dipole senses topological transition in a moiré superlattice, Nature Physics **18**, 765 (2022).

[12] M. Huang et al., Giant nonlinear Hall effect in twisted bilayer $WSe_2$, National Science Review **10**, nwac232 (2023).

[13] L. Min et al., Strong room-temperature bulk nonlinear Hall effect in a spin-valley locked Dirac material, Nature Communications **14**, 364 (2023).

[14] J.-E. Lee et al., Spin-orbit-splitting-driven nonlinear Hall effect in NbIrTe4, Nature Communications **15**, 3971 (2024).

[15] P. He, H. Isobe, D. Zhu, C.-H. Hsu, L. Fu, and H. Yang, Quantum frequency doubling in the topological insulator $Bi_2Se_3$, Nature Communications **12**, 698 (2021).

[16] K. Yasuda, A. Tsukazaki, R. Yoshimi, K. Kondou, K. S. Takahashi, Y. Otani, M. Kawasaki, and Y. Tokura, Current-Nonlinear Hall Effect and Spin-Orbit Torque Magnetization Switching in a Magnetic Topological Insulator, Phys. Rev. Lett. **119**, 137204 (2017).

[17] D. Kaplan, T. Holder, and B. Yan, Unification of Nonlinear Anomalous Hall Effect and Nonreciprocal Magnetoresistance in Metals by the Quantum Geometry, Phys. Rev. Lett. **132**, 026301 (2024).

[18] Y. Gao, S. A. Yang, and Q. Niu, Field Induced Positional Shift of Bloch Electrons and Its Dynamical Implications, Phys. Rev. Lett. **112**, 166601 (2014).

[19] C. Wang, Y. Gao, and D. Xiao, Intrinsic Nonlinear Hall Effect in Antiferromagnetic Tetragonal CuMnAs, Phys. Rev. Lett. **127**, 277201 (2021).

[20] A. Gao et al., Quantum metric nonlinear Hall effect in a topological antiferromagnetic heterostructure, Science **381**, 181 (2023).

[21] N. Wang et al., Quantum-metric-induced nonlinear transport in a topological antiferromagnet, Nature **621**, 487 (2023).

[22] S. Sankar et al., Experimental Evidence for a Berry Curvature Quadrupole in an Antiferromagnet, Phys. Rev. X **14**, 021046 (2024).
[23] S. Lai et al., Third-order nonlinear Hall effect induced by the Berry-connection polarizability tensor, Nature Nanotechnology **16**, 869 (2021).
[24] H. Yu, X. Li, Y.-Q. Bie, L. Yan, L. Zhou, P. Yu, and G. Yang, Quantum metric third-order nonlinear Hall effect in a non-centrosymmetric ferromagnet, Nature Communications **16**, 7698 (2025).
[25] H. Li, C. Zhang, C. Zhou, C. Ma, X. Lei, Z. Jin, H. He, B. Li, K. T. Law, and J. Wang, Quantum geometry quadrupole-induced third-order nonlinear transport in antiferromagnetic topological insulator MnBi2Te4, Nat Commun **15**, 7779 (2024).
[26] C. Wang et al., Room-temperature third-order nonlinear Hall effect in Weyl semimetal $TaIrTe_4$, National Science Review **9**, nwac020 (2022).
[27] X.-G. Ye, H. Liu, P.-F. Zhu, W.-Z. Xu, S. A. Yang, N. Shang, K. Liu, and Z.-M. Liao, Control over Berry Curvature Dipole with Electric Field in WTe 2, Phys. Rev. Lett. **130**, 016301 (2023).
[28] J. Yang, L. Wei, Y. Li, L. Chen, W. Niu, S. Wang, F. Li, P. Liu, S. Zhou, and Y. Pu, Electric field control of nonlinear Hall effect in the type-II Weyl semimetal TaIrTe4, Appl. Phys. Lett. **127**, 033102 (2025).
[29] H. Li, M. Li, R.-C. Xiao, W. Liu, L. Wu, W. Gan, H. Han, X. Tang, C. Zhang, and J. Wang, Current induced second-order nonlinear Hall effect in bulk $WTe_2$, Applied Physics Letters **123**, 163102 (2023).
[30] A. Kofuji, Y. Michishita, and R. Peters, Effects of strong correlations on the nonlinear response in Weyl-Kondo semimetals, Phys. Rev. B **104**, 085151 (2021).
[31] S. Dzsaber et al., Giant spontaneous Hall effect in a nonmagnetic Weyl–Kondo semimetal, Proceedings of the National Academy of Sciences **118**, e2013386118 (2021).
[32] H. (Hugo) Wang et al., Strong electron-phonon coupling driven pseudogap modulation and density-wave fluctuations in a correlated polar metal, Nature Communications **14**, 5769 (2023).
[33] Y. Yuan, P. Kissin, D. Puggioni, K. Cremin, S. Lei, Y. Wang, Z. Mao, J. M. Rondinelli, R. D. Averitt, and V. Gopalan, Ultrafast quasiparticle dynamics in the correlated semimetal $Ca_3Ru_2O_7$, Phys. Rev. B **99**, 155111 (2019).
[34] G. Cao, L. Balicas, Y. Xin, J. E. Crow, and C. S. Nelson, Quantum oscillations, colossal magnetoresistance, and the magnetoelastic interaction in bilayered $Ca_3Ru_2O_7$, Phys. Rev. B **67**, 184405 (2003).
[35] X. N. Lin, Z. X. Zhou, V. Durairaj, P. Schlottmann, and G. Cao, Colossal Magnetoresistance by Avoiding a Ferromagnetic State in the Mott System $Ca_3Ru_2O_7$, Phys. Rev. Lett. **95**, 017203 (2005).
[36] W. Bao, Z. Q. Mao, Z. Qu, and J. W. Lynn, Spin Valve Effect and Magnetoresistivity in Single Crystalline $Ca_3Ru_2O_7$, Phys. Rev. Lett. **100**, 247203 (2008).
[37] Q. Faure, C. D. Dashwood, C. V. Colin, R. D. Johnson, E. Ressouche, G. B. G. Stenning, J. Spratt, D. F. McMorrow, and R. S. Perry, Magnetic structure and field dependence of the cycloid phase mediating the spin reorientation transition in $Ca_3Ru_2O_7$, Phys. Rev. Res. **5**, 013040 (2023).
[38] D. A. Sokolov et al., Metamagnetic texture in a polar antiferromagnet, Nature Physics **15**, 671 (2019).
[39] I. Marković et al., Electronically driven spin-reorientation transition of the correlated polar metal $Ca_3Ru_2O_7$, Proceedings of the National Academy of Sciences **117**, 15524 (2020).
[40] H. Xing et al., Existence of electron and hole pockets and partial gap opening in the correlated semimetal $Ca_3Ru_2O_7$, Phys. Rev. B **97**, 041113 (2018).

[41] M. Horio et al., Electronic reconstruction forming a C2-symmetric Dirac semimetal in Ca3Ru2O7, Npj Quantum Materials **6**, 29 (2021).
[42] Y. Zhao, Z. Mao, and B. Yan, Nonlinear Transport Signatures of Hidden Symmetry Breaking in a Weyl Altermagnet, Phys. Rev. B **112**, 165127 (2025).
[43] A. M. León, J. W. González, and H. Rosner, $Ca_3Ru_2O_7$: Interplay among degrees of freedom and the role of the exchange correlation, Phys. Rev. Mater. **8**, 024411 (2024).
[44] D. Puggioni, M. Horio, J. Chang, and J. M. Rondinelli, Cooperative interactions govern the fermiology of the polar metal $Ca_3Ru_2O_7$, Phys. Rev. Res. **2**, 023141 (2020).
[45] S. Mali et al., Probing hidden symmetry via nonlinear transport in an altermagnet candidate $Ca_3Ru_2O_7$, Nature Communications **17**, 3074 (2026).
[46] A. Saha et al., Evaluating the Structural Basis for Polar Altermagnet Candidate $Ca_3(Ru,Ti)_2O_7$, arXiv Preprint arXiv:2606.16181 (2026).
[47] See Supplemental Material at [URL will be inserted by publisher] for device fabrication, additional measurements on device S2, theoretical calculations, and comparison of third-order nonlinear responsivities, which includes Refs. [57–65].
[48] S. Lei et al., Observation of Quasi-Two-Dimensional Polar Domains and Ferroelastic Switching in a Metal, $Ca_3Ru_2O_7$, Nano Lett. **18**, 3088 (2018).
[49] L. Das et al., Two-carrier Magnetoresistance: Applications to $Ca_3Ru_2O_7$, J. Phys. Soc. Jpn. **90**, 054702 (2021).
[50] C. P. Puls, X. Cai, Y. Zhang, J. Peng, Z. Mao, and Y. Liu, Structural and metal-insulator transitions in ionic liquid-gated $Ca_3Ru_2O_7$ surface, Applied Physics Letters **104**, 253503 (2014).
[51] H. Kishida, H. Matsuzaki, H. Okamoto, T. Manabe, M. Yamashita, Y. Taguchi, and Y. Tokura, Gigantic optical nonlinearity in one-dimensional Mott–Hubbard insulators, Nature **405**, 929 (2000).
[52] H. Liu, Y. Li, Y. S. You, S. Ghimire, T. F. Heinz, and D. A. Reis, High-harmonic generation from an atomically thin semiconductor, Nature Physics **13**, 262 (2017).
[53] R. E. F. Silva, I. V. Blinov, A. N. Rubtsov, O. Smirnova, and M. Ivanov, High-harmonic spectroscopy of ultrafast many-body dynamics in strongly correlated systems, Nature Photonics **12**, 266 (2018).
[54] Y. Murakami, M. Eckstein, and P. Werner, High-Harmonic Generation in Mott Insulators, Phys. Rev. Lett. **121**, 057405 (2018).
[55] H. Jiang et al., Probing interplay of topological properties and electron correlation in TaIrTe4 via nonlinear Hall effect, Nat Commun **16**, 6351 (2025).
[56] R. Y. Chu et al., Third-Order Nonlinear Hall Effect in Altermagnet $RuO_2$, Phys. Rev. Lett. **135**, 216703 (2025).
[57] H. Shi et al., Magnetic field tunable third-order nonlinear transport in a heterodimensional superlattice, Phys. Rev. B **111**, 235434 (2025).
[58] Z.-H. Chen et al., Charge density wave modulated third-order nonlinear Hall effect in 1T-$VSe_2$ nanosheets, Phys. Rev. B **110**, 235135 (2024).
[59] T.-Y. Zhao et al., Gate-Tunable Berry Curvature Dipole Polarizability in Dirac Semimetal $Cd_3As_2$, Phys. Rev. Lett. **131**, 186302 (2023).
[60] S. Li et al., Giant Third-Order Nonlinear Hall Effect in Misfit Layer Compound $(SnS)_{1.17}(NbS_2)_3$, ACS Appl. Mater. Interfaces **16**, 11043 (2024).
[61] G. Kresse and J. Furthmüller, Efficient iterative schemes for ab initio total-energy calculations using a plane-wave basis set, Phys. Rev. B **54**, 11169 (1996).
[62] J. P. Perdew et al., Restoring the Density-Gradient Expansion for Exchange in Solids and Surfaces, Phys. Rev. Lett. **100**, 136406 (2008).

[63] S. L. Dudarev et al., Electron-energy-loss spectra and the structural stability of nickel oxide: An LSDA+U study, Phys. Rev. B **57**, 1505 (1998).
[64] Y. Yoshida et al., Crystal and magnetic structure of $Ca_3Ru_2O_7$, Phys. Rev. B **72**, 054412 (2005).
[65] I. Souza, N. Marzari, and D. Vanderbilt, Maximally localized Wannier functions for entangled energy bands, Phys. Rev. B **65**, 035109 (2001).